\documentclass[prl,superscriptaddress,preprintnumbers,twocolumn,longbibliography]{revtex4-1} 
\usepackage{amsmath}
\usepackage{amssymb}
\usepackage{amsthm}
\usepackage{amsfonts}
\usepackage{algorithmic}
\usepackage{enumerate}
\usepackage{latexsym}
\usepackage{graphicx}
\usepackage{bm}
\usepackage{subfigure}
\usepackage{setspace}
\usepackage[dvips]{color}
\usepackage{hyperref}
\usepackage{bm}
\usepackage{mathrsfs}
\begin{document}

\title{Effects of Boundary on Orbital Magnetization for a Bilayer System with Different Chern Numbers}

\author{Si-Si Wang}
\affiliation{School of Physics and Materials Science, Guangzhou University, 510006 Guangzhou, China}
\affiliation{School of Mathematics and Information Science, Guangzhou University, 510006 Guangzhou, China}
\affiliation{Huangpu Research and Graduate School of Guangzhou University, 510700 Guangzhou, China}

\author{Yan Yu}
\affiliation{SKLSM, Institute of Semiconductors, Chinese Academy of Sciences, P.O. Box 912, Beijing 100083, China}
\affiliation{School of Physical Sciences, University of Chinese Academy of Sciences, Beijing 100049, China}

\author{Ji-Huan Guan}
\affiliation{Beijing Academy of Quantum Information Sciences, Beijing 100193, China}
\affiliation{SKLSM, Institute of Semiconductors, Chinese Academy of Sciences, P.O. Box 912, Beijing 100083, China}

\author{Yi-Ming Dai}
\affiliation{School of Physics and Materials Science, Guangzhou University, 510006 Guangzhou, China}

\author{Hui-Hui Wang}
\affiliation{School of Physics and Materials Science, Guangzhou University, 510006 Guangzhou, China}
\affiliation{Huangpu Research and Graduate School of Guangzhou University, 510700 Guangzhou, China}

\author{Yan-Yang Zhang}
\email{yanyang@gzhu.edu.cn}
\affiliation{School of Physics and Materials Science, Guangzhou University, 510006 Guangzhou, China}
\affiliation{Huangpu Research and Graduate School of Guangzhou University, 510700 Guangzhou, China}
\affiliation{School of Mathematics and Information Science, Guangzhou University, 510006 Guangzhou, China}

\date{\today}

\begin{abstract}
The real space formalism of orbital magnetization (OM) is an average of the local OM over some appropriate region of the system.
Previous studies prefer a bulk average (i.e., without including boundaries). Based on a bilayer model with an adjustable Chern number at half filling, we numerically investigate the effects from boundaries on the real space expressions of OM. The size convergence processes of its three constituent terms $M_{\mathrm{LC}}$, $M_{\mathrm{IC}}$, $M_{\mathrm{BC}}$ are analysed. The topological term $M_{\mathrm{BC}}$ makes a nonnegligible contribution from boundaries as a manifestation of edge states, especially in the case of nonzero Chern numbers. However, we show that the influence of the boundary on $M_{\mathrm{LC}}$ and $M_{\mathrm{IC}}$ exactly compensates that on $M_{\mathrm{BC}}$. This compensation effect leads to the conclusion that the whole sample average is also a correct algorithm in the thermodynamic limit, which gives the same value as those from the bulk average and the $k$ space formula. This clarification will be beneficial to further studies on orbitronics, as well as the orbital magnetoelectric effects in higher dimensions.
	
\end{abstract}

\maketitle

\section{I. Introduction}
The magnetization of solids comes from two parts: spin magnetization and orbital magnetization (OM). For the OM, the relevant theory had been confined to calculating the change of OM as a bulk quantity\cite{Mauri1996,MauriF1996,Pickard2002,Sebastiani2002}. Until 2005, the modern theory of OM in solids was developed\cite{Vanderbilt2018}. The OM is then found to be closely related to topological properties of solids\cite{Ceresoli2006,Malashevich2010,Bianco2011,Bianco2013,XiDai2019,JPHanke2016,Lux2018,Helin2020}. Recently it is proposed as the central quantity of a topological memory device\cite{TopologicalMemory}. How to tune the OM in solids is an important objective of orbitronics\cite{Go2017,Bhowal2020,DGo2021}. Besides the induction of OM from external fields\cite{Dong2020,Xiao2021,Hamada2020,Sharpe2019,Serlin2020,Su2020,He2020,Huang2021}, another method is to produce a spontaneous OM, such as, by engineering material structures\cite{Cinal2022}. These potential applications raise demands of predicting materials' OM based on a deeper understanding of calculation methods.

Similar to Chern number as the integration of the Berry curvature in the momentum space ($k$ space), the modern theory also expresses OM as a $k$ space integral, which can be derived from the semi-classical method\cite{Xiao2005,Xiao2010}, the Wannier representation\cite{Thonhauser2005,Resta2005,Thonhauser2011}, or the perturbation theory\cite{Shi2007}. This expression consists of two terms associated with the local circulation (LC) and the itinerant circulation (IC) motions respectively. With one additional term including the contribution from Berry curvature (BC), it addresses on the same ground the normal insulators, Chern insulators, and metals\cite{Ceresoli2006}.

Afterwards, there is another formalism of expressing OM and Chern number for a bounded sample, which define them as local quantities in the real space ($r$ space)\cite{Bianco2011,Bianco2013,Bianco2016,Drigo2020,Sykes2021}. In the thermodynamic limit, an average of these local quantities over an appropriate region gives correct values consistent with the $k$ space formalism\cite{Bianco2013,MarrazzoResta2017}. Furthermore, the $r$ space formalism of the orbital magnetoelectric coupling, i.e., the linear response of OM respect to an electric field, has also been developed\cite{Malashevich2010,Olsen2017,Varnava2018,Varnava2020}, which paves route towards deep understanding of axion physics in solids\cite{AxionReview1,AxionReview2}.
While the $k$ space theory requires translation symmetry of solids, the $r$ space method is suitable for a system with arbitrary boundary conditions, or with spatial inhomogeneity.

It has been recently realized that these $r$ space descriptions are not only mathematical structures but also with physical meanings. Around the topological phase transition, numerical simulations show that the local Chern number plays the role as that of the order parameter in thermodynamic phase transitions, with familiar properties such as universal scaling and divergence of correlation length\cite{Caio2019,Ulcakar2020}.
In twisted bilayer graphene, a large but non-uniformly distributed local OM
were theoretical proposed\cite{XiDai2019}, which were experimentally detected\cite{OMBilayerGraphene}.

Nevertheless in this $r$ space formalism, there is still some vagueness on the effects from boundaries. For the calculation of Chern number, the average should be over the bulk region \emph{only}, and the average over the whole sample (bulk plus boundaries) is always zero even for a Chern insulator\cite{Bianco2011,Bianco2013}. Therefore the contribution from boundaries appears to be nonnegligible but ``unphysical''. For the calculation of OM for a normal insulator, however, the average can be over the bulk \emph{or} the whole sample, with the former a faster convergence\cite{Bianco2013}. Another option is to redefine a boundary-determined term to the OM in the $k$ space formalism\cite{Bianco2016}. Back to the $r$ space, the whole sample average of OM for a Chern insulator seems to be questionable due to the BC term closely related to Chern number, for which the boundary contribution seems to be unphysical as mentioned above. As for the $r$ space formalism of the orbital magnetoelectric coupling, it is suggested that the average should be over the whole sample with the same footing, even with nontrivial topology\cite{Olsen2017}. Therefore, a detailed understanding of boundary influence on OM is not very satisfactory so far. Is it really necessary to or not to include boundaries on OM when one is calculating in the $r$ space?

In this paper, we revisit these problems of the $r$ space formalism numerically, by scrutinizing individual contributions to OM from the bulk and boundary regions, as well as individual contributions from the LC, IC and BC terms, respectively. Our calculations are based on a bilayer system with an adjustable Chern number. Compared to a monolayer one, it has better tunability of properties (symmetry, band structure and Chern number), and it is also a good starting point for future investigations of higher dimensions, e.g., the axion orbital effects in three dimensions\cite{Olsen2017,Varnava2018,Varnava2020,Wang2019}. We confirm that, the result of a whole sample average is still consistent with the $k$ space method in the thermodynamic limit even for Chern insulators. We also find that, contributions from different origins, i.e., bulk versus boundary, LC, IC versus BC terms, often carry different signs and cancel each other to result in a small value of total OM. Therefore in the materials engineering, a possible means to enhance the OM is to suppress one of them.

This paper is organized as follows. In Sec.II we present the Hamiltonian. In Sec. III we lay out the $r$ space method and the $k$ space method of OM. In Sec. IV and V we study the influence of sample boundary on OM. Some small size effects are discussed in Section VI. Conclusions are drawn in Sec. VII.

\section{II. The Model}
\begin{figure}[htbp]
	\includegraphics*[width=0.45\textwidth]{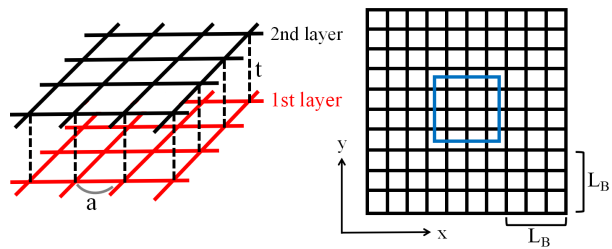}
	\caption{ The left panel denotes the lattice structure of our constructed bilayer model. The first layer(Red) and the second layer (black) have different model parameters and the same lattice constant $a$. $t$ is the inter-layer coupling strength. The right panel is the top view of our model. The region enclosed by blue lines represents the bulk region. $L_{B}$ denotes the thickness of the boundaries. Here $L_{B}=4$. }
	\label{lattice}
\end{figure}

The lattice structure of our model is composed of two square-lattice layers extending in the $x$-$y$ plane as illustrated in Fig. \ref{lattice}. The first layer (red) and the second layer (black) are stacked along the $z$ direction.
Each layer is a quantum anomalous Hall (QAH) model with a tunable Chern number and there exist inter-layer couplings. Experimentally, such topological bilayers can be realized by using state-of-the-art
technologies of micro hetero-structures\cite{Fabrication1,Fabrication2,Fabrication3},
ultracold atomic\cite{FabricationOL1,FabricationOL2,FabricationOL3}
or photonic systems\cite{FabricationPhot}.

\begin{figure}[htbp]
	\includegraphics*[width=0.45\textwidth]{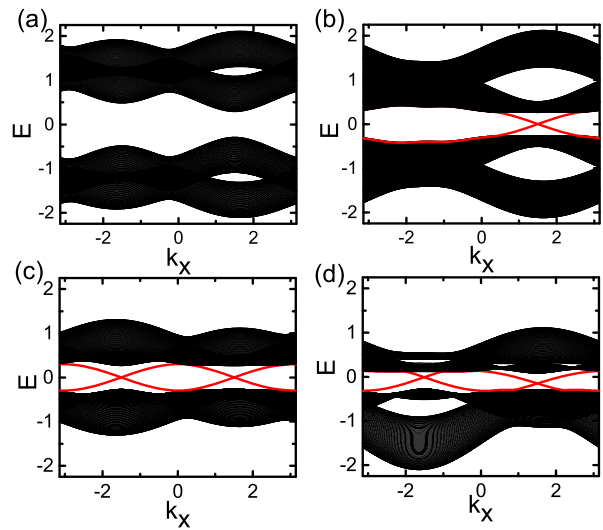}
	\caption{ The band structure of the bilayer QAH model in the quasi-one-dimensional ribbon geometry with width $Ny=80$. Red lines are edge states. The four panels correspond to different parameter settings. (a) $D_1=0$, $M_1=-0.3$, $M_2=-0.5$. (b) $D_1=0$, $M_1=0.3$, $M_2=-0.5$. (c) $D_1=0$, $M_1=0.3$, $M_2=0.5$. (d) $D_1=0.1$, $M_1=0.3$, $M_2=0.5$. The other model parameters are identical: $A_1=A_2=0.3$, $B_1=B_2=0.2$, $D_2=0$, $t=0.1$. Particle-hole symmetry is preserved in panel(a)-(c), and is broken in panel(d). }
	\label{FigDispersion}
\end{figure}

The Hamiltonian of this bilayer system can be generally expressed as\cite{Wang2019}
\begin{equation}
	\mathcal{H}_{\mathrm{bi}}=\mathcal{H}_1+\mathcal{H}_2+\mathcal{H}_c,\label{EqTotalHamiltonian}
\end{equation}
where $\mathcal{H}_L$ ($L=1,2$) is the Hamiltonian for the $L$-th layer, and
$\mathcal{H}_c$ represents the coupling between them. Within each layer, we choose the spin up component of the Bernevig-Hughes-Zhang model\cite{Bernevig2006} to construct a spinless QAH model. With one $s$ orbital and one $p$ orbital on each site, this is one of the minimum models of the QAH state. In the $k$ space, the $L$-th layer Hamiltonian reads
\begin{equation}
	\mathcal{H}_L=\sum_{\bm{k,\alpha\beta}}c^{\dagger}_{L;\bm{k}\alpha}H_{L;\alpha\beta}(\bm{k}) c_{L;\bm{k}\beta},\label{EqHL0}
\end{equation}
where $c^{\dagger}_{L;\bm{k}\alpha}$ ($c_{L;\bm{k}\alpha}$) creates (annihilate) an electron with wavenumber $\bm{k}$ and orbital $\alpha\in \{s,p\}$ in layer $L\in\{1,2\}$. Here, $H_{L;\alpha\beta}(\bm{k})$ is a $2 \times 2$ matrix as\cite{Bernevig2006}
\begin{eqnarray} \label{EqH}
	H_L(\bm{k})&=&\varepsilon_L(\bm{k})I_{2\times 2}+\sum_{i}d^i_L(\bm{k})\sigma_i \label{EqBHZ}\\
	\varepsilon_L(\bm{k})&=&-2D_L\big[2-\cos ( k_{x}-q_L^x )-\cos ( k_{y}-q_L^y )\big] \nonumber \\
	d^1_L(\bm{k})&=&A_L\sin ( k_{x}-q_L^x ), \quad d^2_L(\bm{k})=A_L\sin ( k_{y}-q_L^y )\nonumber\\
	d^3_L(\bm{k})&=&M_L-2B_L\big[2-\cos (k_{x}-q_L^x) -\cos ( k_{y}-q_L^y)\big],\nonumber
\end{eqnarray}
where $\sigma_i$ are the Pauli matrices acting on the orbital space, and $\bm{q}_L=(q_{x},q_{y})$ represents the momentum shift of the band structure in the Brillouin zone for the $L$-th layer which helps us to visually distinguish two groups of edge states (originated from two layers) as Fig. (\ref{FigDispersion}) shown.
The real space version of the $L$-th layer Hamiltonian $\mathcal{H}_L=\sum_{ij,\alpha\beta}c^{\dagger}_{L;i\alpha}H_{L;\alpha\beta}(i,j) c_{L;j\beta}$ can be obtained from Eqs. (\ref{EqHL0}) and (\ref{EqBHZ})
by performing a straightforward inverse Fourier transformation
$c_{L;\bm{k}\beta }=\frac{1}{\sqrt{V}}\sum_{i}c_{L;i\beta }e^{-i\bm{k}\cdot {\bm{r}_{i}}}$, where $i$ is the site index.

In the absence of the inter-layer coupling $H_c$, the band structure and Chern number of
each layer can be tuned independently by varying model parameters.
For example, the band gap is $2|M_L|$, and the Chern number
\begin{equation}
	C_L=
	\left\{
	\begin{array}{lll}
		+1,&\quad 0<M_L/2B_L<2 \\
		0,&\quad M_L/2B_L<0\\
		-1,&\quad . 
	\end{array}
	\right.
\end{equation}
The inter-layer coupling is considered to be the simple
form in real space as
\begin{equation}
	\mathcal{H}_c=\sum_{i\alpha}\big(t c^{\dagger}_{1;i\alpha}c_{2;i\alpha}+\mathrm{H. c.}\big),
	\label{EqHc}
\end{equation}
where $t$ is the inter-layer coupling strength.
If the adiabatic tuning of the inter-layer coupling term (\ref{EqHc}) does not close the bulk gap, the Chern number of the bilayer system is just the sum of those of each layer, $C=C_1+C_2$.
Hence, it is no surprise that the Chern number of the system changes with the band gap $|M_L|$ and the inter-layer coupling strength $t$\cite{Wang2019}.

In Fig. \ref{FigDispersion}, we present band structures of the quasi-one-dimensional (quasi-1D) ribbon with several typical settings of model parameters.
The number of edge states (red curves) in the bulk band gap reflects the total Chern number at half filling. For example, there is one group of edge states in Fig. \ref{FigDispersion} (b) so that the Chern number of the system is $1$. Similarly, the Chern numbers in panel (a), (c) and (d) are $C=0$, $C=2$ and $C=2$ respectively.  From the parameter $q_L^x$, it can be deduced that the edge state on the left ($k_{x}<0$) originates from the second layer, and the other pair originates from the first layer. The model parameters $D_L$ determine the e-h symmetry. Only when both of $D_{L}$($L\in\{1,2\}$) are zero, the band structure is symmetric respect to energy $E=0$. This symmetry is broken in Fig. \ref{FigDispersion} (d) with $D_1=0.1$.

\section{III. Methods}
In the modern theory, the macroscopic OM of a general two dimensional crystalline solid can be expressed as an integration over the Brillouin zone\cite{Thonhauser2005,Resta2005,Ceresoli2006},
\begin{equation}\label{EqMk}
\begin{aligned}
 M&=M_{\mathrm{LC}}+M_{\mathrm{IC}}+M_{\mathrm{BC}},\\
 M_{\mathrm{LC}}&=\frac{e}{2\hbar c}\mathrm{Im}\sum_n\int_{\varepsilon_{nk}\leqslant\mu}
 \frac{\mathrm{d}\bm{k}}{(2\pi)^2}\langle\partial_{\bm{k}}u_{n\bm{k}}|
 \times H_{\bm{k}} |\partial_{\bm{k}}u_{n\bm{k}}\rangle \\
 M_{\mathrm{IC}}&=\frac{e}{2\hbar c}\mathrm{Im}\sum_n\int_{\varepsilon_{nk}\leqslant\mu}
 \frac{\mathrm{d}\bm{k}}{(2\pi)^2}\varepsilon_{n\bm{k}}\langle\partial_{\bm{k}}u_{n\bm{k}}|
 \times|\partial_{\bm{k}}u_{n\bm{k}}\rangle \\
 M_{\mathrm{BC}}&=-\mu \frac{e}{2\pi\hbar c}\mathcal{C}\\
 \mathcal{C}&=\frac{1}{2\pi}\mathrm{Im}\sum_n\int_{\varepsilon_{nk}\leqslant\mu}
 \mathrm{d}\bm{k}\langle\partial_{\bm{k}}u_{n\bm{k}}|
 \times|\partial_{\bm{k}}u_{n\bm{k}}\rangle
\end{aligned}
\end{equation}
where $|u_{n\bm{k}}\rangle$ is the cell-periodic Bloch function, and $\varepsilon_{nk}$ is the Bloch eigenvalue which satisfies $H_{\bm{k}}|u_{n\bm{k}}\rangle=\varepsilon_{nk}|u_{n\bm{k}}\rangle$. All summations in equation (\ref{EqMk}) are over occupied bands $n$ up to $\mu$. Three constituent terms of $M$ correspond to the local circulation (LC), the itinerant circulation (IC), and the Berry curvature (BC) contributions respectively. The dimensionless number $\mathcal{C}$ is quantized as the Chern number $C$ (the topological invariant) when $\mu$ is in the bulk gap, but may not be quantized if $\mu$ is not in the gap.
In numerical calculations, partial derivatives in the above equations cannot be simply replaced by finite differences. Instead, a fixed gauge must be guaranteed along neighboring mesh points. Details of this algorithm is introduced in the Appendix.

Analogous to the above $k$ space integration, OM and its constituent terms can also be expressed as a real space integration as\cite{Bianco2013,MarrazzoResta2017}
\begin{equation}\label{EqMr-local}
	\begin{aligned}
		M_{\alpha}&=\frac{1}{A}\int_{A} d{\bm{r}}\;m_{\alpha}\left( \bm{r}\right),\quad \alpha\in\{\mathrm{LC, \; IC, \;  BC}\} \\
        m\left( \bm{r}\right)&={m_{\mathrm{LC}}\left( \bm{r}\right)}+{m_{\mathrm{IC}}\left( \bm{r}\right)}+{m_{\mathrm{BC}}\left( \bm{r}\right)}\\
		{m_{\mathrm{LC}}\left( \bm{r}\right)}&=\frac{e}{\hbar c}\mathrm{Im}\langle \bm{r}| PxQHQyP|\bm{r}\rangle\\
	    {m_{\mathrm{IC}}\left( \bm{r}\right)}&=-\frac{e}{\hbar c}\mathrm{Im}\langle \bm{r}| QxPHPyQ|\bm{r}\rangle \\
        {m_{\mathrm{BC}}\left( \bm{r}\right)}&=-\mu\frac{e}{2\pi\hbar\ c}\emph{c}\left( \bm{r}\right) \\
		{\emph{c}\left( \bm{r}\right)}&=4\pi \mathrm{Im}\langle \bm{r}|QxPyQ|\bm{r}\rangle \\
	\end{aligned}
\end{equation}
where the ground-state projector $P$ and its complement $Q$ satisfy  $Q=1-P$ (see Appendix for details). Here $m_{\mathrm{LC}}\left( \bm{r}\right)$, $m_{\mathrm{IC}}\left( \bm{r}\right)$ and $m_{\mathrm{BC}}\left( \bm{r}\right)$ correspond to the local markers of the corresponding terms in Eq.(\ref{EqMk}). The local Chern number $\emph{c}\left( \bm{r}\right)$ is also called the topological marker, whose real space distribution has attracted some attentions\cite{Caio2019,Ulcakar2020}.

Now the total OM $M$ and its three constituent terms $M_{\alpha}$ have the form of a real space average over some region with an area $A$, i.e.,
\begin{equation}
\frac{1}{A}\int_{A} d{\bm{r}}\langle\cdots\rangle.\nonumber
\end{equation}
This region can be the whole sample, or just the bulk region around the sample center (i.e., excluding the boundary region), as illustrated in Fig. \ref{lattice} (b). As for the Chern number, it is worth emphasizing that $\emph{c}\left( \bm{r}\right)$ have opposite signs in the boundary and bulk regions: $\emph{c}\left( \bm{r}\right)$ in the bulk region are equal to the Chern number $C$, while they have opposite values in the boundary region, so that its average over the whole sample always vanish\cite{Bianco2011,Bianco2013}. This is related to the mathematical fact of a projector over a finite-dimensional manifold\cite{Bianco2011}.

Hence for calculating the OM by using Eq. (\ref{EqMr-local}), this feature has no effect on a normal insulator about whether the trace (or integration) should be taken over the whole sample or the bulk region, but it is different for Chern insulators due to the existence of the BC term. Previous works indicate that the trace of $\mathit{m}\left( \bm{r}\right)$ should also be taken over the bulk region for Chern insulators\cite{Bianco2013}. Nevertheless, a remaining question is whether there is any underlying physical meaning for ignoring boundaries. To figure this out, it is necessary to study the influence of the sample boundary on $M$, especially how the boundary affect the other two terms of $M$, namely, $M_{\mathrm{IC}}$ and $M_{\mathrm{LC}}$.

In the following, we will investigate the influence of boundaries on $M$ and its three constituent terms $M_{\alpha}$,
and check whether the average can be taken over the whole sample or not.
This problem is important for the calculation of $M$ itself, especially for samples with inhomogeneity or with small sizes.
Furthermore, it will offer insights on the real space algorithm of axion magnetoelectric coupling\cite{Wang2019}.

\section{IV. Normal Insulator}

\begin{figure}[htbp]
	\includegraphics*[width=0.48\textwidth]{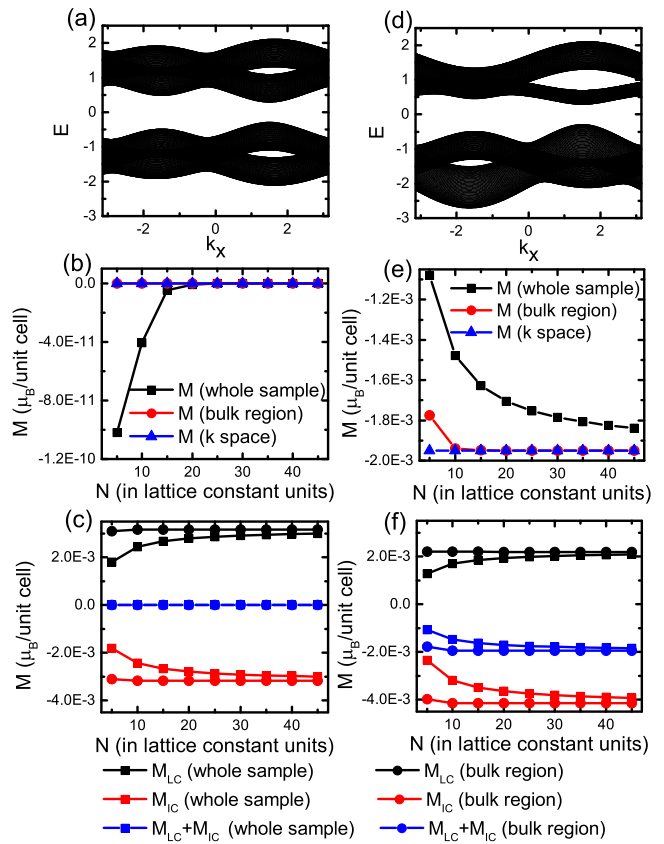}
\caption{ Bilayer with $C=0$. Upper row: band structure of a quasi-1D ribbon. Middle row: total OM as a function of sample size, calculated from three methods. Lower row: $M_{\mathrm{LC}}$(black lines) and $M_{\mathrm{IC}}$(red lines) as functions of the sample size $N$. The left (right) column corresponds to the case with (without) e-h symmetry, by setting $D_1=D_2=0$ ($D_1=0.1,D_2=0$). Square: OM from the whole sample average. Circle: from bulk region. Up triangle: from $k$ space method. Other model parameters are identical for all panels: $A_1=A_2=0.3$, $B_1=B_2=0.2$, $M_1=-0.3$, $M_2=-0.5$, $t=0.1$. The Fermi energy $\mu=0.05$.	}
\label{FigM-c=0}
\end{figure}

Let us first focus on the topologically trivial
insulator as shown in Fig. \ref{FigM-c=0}. The left (right) column of Fig. \ref{FigM-c=0} corresponds to the case of e-h symmetric (asymmetric) band structure respect to $E=0$.
To see the convergence of the $r$ space method with increasing sample size $N$ (corresponding to a bilayer with $2N^2$ lattice sites), we plot corresponding OMs calculated from an average over the whole sample (black line with square dots) and the bulk region (red line with circle dots) in Fig. \ref{FigM-c=0} (b) and \ref{FigM-c=0} (e). Here, we fix $\mu=0.05$ in the band gap. As illustrated in Fig. \ref{lattice}, the bulk region refers to the inner part of the sample enclosed by the blue dashed square, while the rest part is the boundary region. OMs from the $k$ space method is also shown as a comparison. In the following, we select the central quarter ($\frac{N}{2}\times\frac{N}{2}$) of the sample as the bulk region, unless otherwise specified (e.g., in Section VI). The thickness of the boundaries is $L_{B}=\frac{N}{4}$.

One can see that $M$ calculated from $r$ space methods (black and red line) converge to the thermodynamic limit obtained by $k$ space method (blue line), and the average of $\mathit{m}\left( \bm{r}\right)$ over bulk region converges faster than over the whole sample, which is consistent with previous studies\cite{Bianco2013}. In addition, the value of $|M|$ is almost zero when the system satisfies e-h symmetry in Fig. \ref{FigM-c=0}(b), while the value of $|M|$ is greatly increased by breaking e-h symmetry in Fig. \ref{FigM-c=0} (e).

To figure out the reason for the increase of $|M|$ as the e-h symmetry is broken, we further plot its constituent terms, namely $M_{\mathrm{LC}}$ and $M_{\mathrm{IC}}$ as functions of sample size $N$ in Fig. \ref{FigM-c=0} (c) and \ref{FigM-c=0} (f). Notice now the term $M_{\mathrm{BC}}$=0. In Fig. \ref{FigM-c=0} (c), $M_{\mathrm{LC}}$ and $M_{\mathrm{IC}}$ are perfectly opposite to each other, $M_{\mathrm{LC}}=-M_{\mathrm{IC}}$, resulting in a vanishing $M$. This can be understood from Eq. (\ref{EqMrApp}) in Appendix. When the system satisfies e-h symmetry at half filling, the ground state projector $P$ becomes completely equivalent to $Q$, and then $\mathrm{Tr}\left\lbrace PxQHQyP\right\rbrace=\mathrm{Tr}\left\lbrace QxPHPyQ\right\rbrace$. So, the expressions of $M_{\mathrm{LC}}$ and $M_{\mathrm{IC}}$ only differ by a minus sign. This perfect cancelation does not hold when the e-h symmetry is broken, since $P$ and $Q$ are no longer equivalent, as can be seen in Fig. \ref{FigM-c=0}(f). Hence, breaking e-h is an effective method to enhance the OM at half filling.

\begin{figure}[htbp]
   \includegraphics*[width=0.48\textwidth]{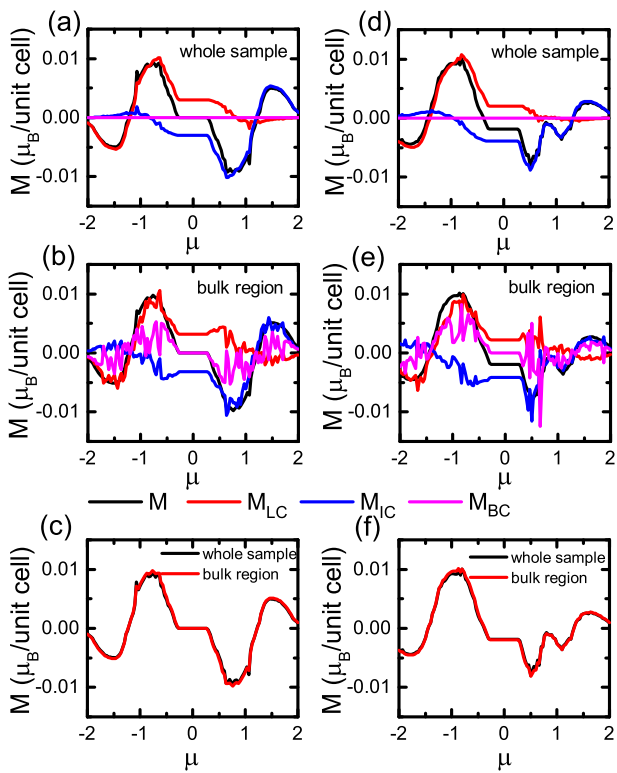}
   \caption{ The OM $M$ and its constituents $M_{\mathrm{LC}}$, $M_{\mathrm{IC}}$ and $M_{\mathrm{BC}}$ as functions of the Fermi energy $\mu$. Upper row: OMs from the whole sample average. Middle row: OMs from the bulk region average. Lower row: comparison of total OM from two methods of real space average. The left (right) column corresponds to case with (without) e-h symmetry by setting $D_1=D_2=0$ ($D_1=0.1,D_2=0$). The other model parameters are the same as Fig. \ref{FigM-c=0}. The sample size $N=40$.}
\label{FigM-EF}
\end{figure}

If the Fermi energy (chemical potential) $\mu$ is moved into the the valence or conductance band, the system will be metallic. Now the expression of $M$ has an extra term $M_{\mathrm{BC}}$ depending on $\mu$ in Eq. (\ref{EqMk}) or (\ref{EqMr-local}), which was zero when $\mu$ is in the band gap. In Fig. \ref{FigM-EF}, we present the Fermi energy dependence of OMs, for the e-h symmetric (left column) and asymmetric (right column) cases, respectively.  We find that
in both cases, $|M|$ increases rapidly once $\mu$ enters either band until an obvious peak, as shown in Fig. \ref{FigM-EF} (c) and (f). To see the origins of the OM, in Fig. \ref{FigM-EF}, we still display three constituent terms $M_{\mathrm{LC}}$, $M_{\mathrm{IC}}$ and $M_{\mathrm{IC}}$, averaged over the whole sample (first row) and over the bulk region (second row). Most contributions are from $M_{\mathrm{LC}}$ and $M_{\mathrm{IC}}$ (which will not be cancelled out due to the inequivalence between projectors $P$ and $Q$), but their weights may be different in different energy regions. For example, as can be seen from panels (a) and (d),
the dominating contribution to the conduction (valence) band peak is from $M_{\mathrm{IC}}$ ($M_{\mathrm{LC}}$).

More details can be noticed by comparing Fig. \ref{FigM-EF} (a) between (b), and (d) between (e). Concrete values of $M_{\mathrm{LC}}$, $M_{\mathrm{IC}}$ and $M_{\mathrm{IC}}$ may be different, when the average is over the whole sample (first row) or over the bulk region (second row). However, their sums, the total OM from both algorithms, matches perfectly at all Fermi energies (last row). This preliminarily indicates that, even with nonzero contribution from $M_{\mathrm{BC}}$, the boundary can safely be included in the calculation of $M$ when using the real space method. We further verify that this is the case for Chern insulators.

\section{V. Chern Insulator}
\subsection{A. Chern number $C=1$}

\begin{figure}[htbp]
	\includegraphics*[width=0.48\textwidth]{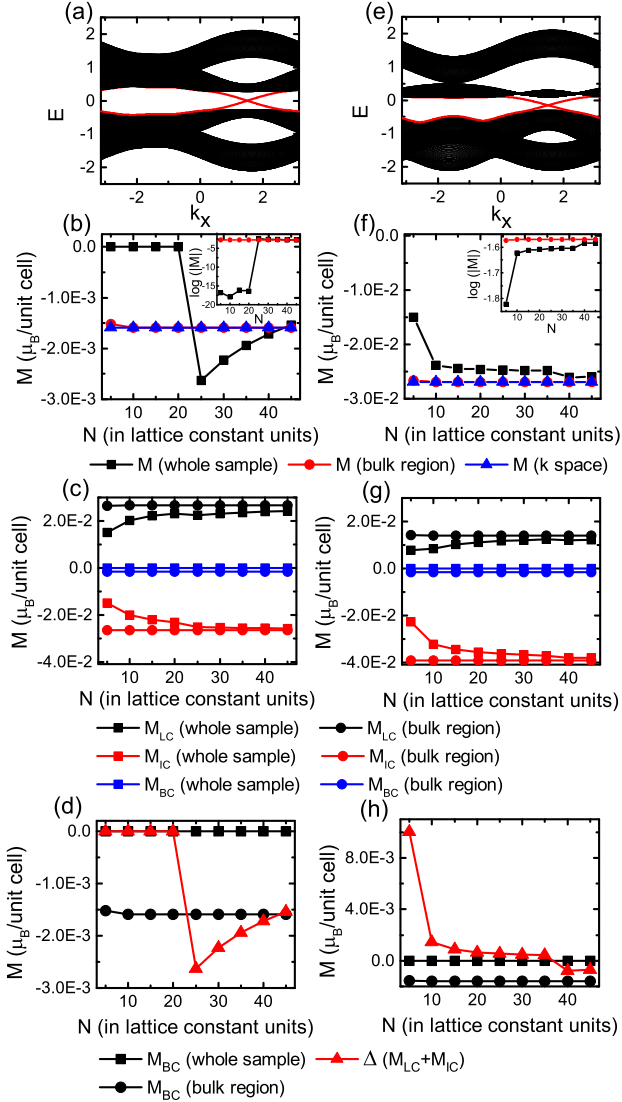}
	\caption{ Bilayer with $C=1$. First row: band structures of a quasi-1D ribbon. Second row: the total OM as a function of the sample size $N$ calculated from different methods: whole sample average (black), bulk region average (red), and $k$ space method (blue). Third row: $M_{\mathrm{LC}}$ (black lines) ,$M_{\mathrm{IC}}$ (red lines) and $M_{\mathrm{BC}}$ (blue lines) as functions of the sample size $N$. Fourth row: The comparison of $M_{\mathrm{BC}}$ (black line) with $\Delta(M_{\mathrm{LC}}+M_{\mathrm{IC}})$ (red line) as functions of the sample size $N$. The left (right) column corresponds to the case with (without) e-h symmetry. Square: whole sample average. Circle: bulk region average. Up triangle: $k$ space method. $M_1=0.3$, and other model parameters are identical to Fig. \ref{FigM-c=0}.}
	\label{FigM-c=1}
\end{figure}

In this part, we focus on the topologically non-trivial phase of the bilayer system. Let us start from the Chern number $C=1$ case with band structures shown in Fig. \ref{FigM-c=1} (a) (e-h symmetric) and \ref{FigM-c=1} (e) (e-h asymmetric). There is only one group of edge states (red lines) originating from the first layer of the system.

We set the Fermi energy $\mu=0.05$ well inside the bulk gap, and calculate $M$ and its local distribution by using Eqs. Eq. (\ref{EqMk}) and (\ref{EqMr-local}). First, let us still check their convergence with increasing size $N$. The second row of Fig. \ref{FigM-c=1} show comparisons of $M$ averaged over the whole sample, the bulk, and from the $k$ space method. Again, all three results agree in the thermodynamic limit, even though the system is topologically non-trivial. This is different from previous conclusions which indicate that the real space average should be taken over the bulk region only for Chern insulators\cite{Bianco2013}.

\begin{figure}[htbp]
	\includegraphics*[width=0.48\textwidth,bb=31 24 482 209]{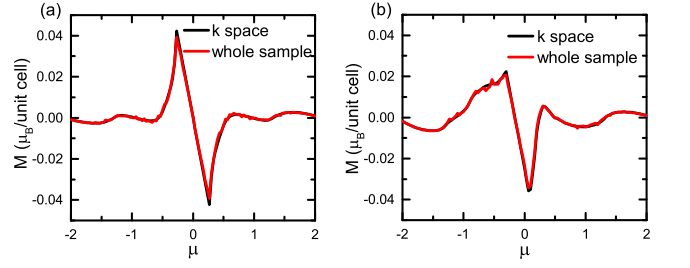}
	\caption{ The total OM as a function of the Fermi energy $\mu$ for the bilayer with $C=1$, calculated from whole sample average (red) and $k$ space method (black). (a) With e-h symmetry. (b) Without e-h symmetry. Model parameters are identical to Fig. \ref{FigM-c=1}. }
	\label{FigM-k&r(c=1)}
\end{figure}

This can be attributed to the real space distribution of local Chern number $c\left( \bm{r}\right)$. Only when averaged over the bulk region of the sample, does the dimensionless number $\mathcal{C}$ equal to the Chern number $C$.
However, when we take the average of $c\left( \bm{r}\right)$ over the whole sample, $\mathcal{C}$ will always vanish, and therefore the contribution from the corresponding term $M_{\mathrm{BC}}=-\mu \frac{e}{2\pi\hbar c}\mathcal{C}$ remains zero.

To further clarify this convergence, we turn to details of three constituent parts, $M_{\mathrm{LC}}$ and $M_{\mathrm{IC}}$ and $M_{\mathrm{BC}}$, presented in the third row of Fig. \ref{FigM-c=1}.
In previous case of normal insulator, we have shown that in the special case of e-h symmetry, $M_{\mathrm{LC}}$ and $M_{\mathrm{IC}}$ at half filling are opposite to each other. We find that this is also the case here with $C=1$, but only when the average is over the bulk region. As displayed in Fig. \ref{FigM-c=1}(c), when averaged over bulk, $M_{\mathrm{LC}}$ (black circle line) and $M_{\mathrm{IC}}$ (red circle line) are symmetric with respect to zero. But when the average is over the whole sample, they are no longer symmetric. Meanwhile, $M_{\mathrm{BC}}$ changes from a finite value (blue circle line) to zero (blue square line) as the average changes from bulk region to the whole sample. This indicates that the existence of boundary states not only affects the geometric term $M_{\mathrm{BC}}$, but also affects two non-geometric terms, $M_{\mathrm{LC}}$ and $M_{\mathrm{IC}}$.

Now let us scrutinize the effect of boundaries. For these two types of real space averages of OM, i.e., over the whole sample and over the bulk region, their difference is that of the boundary. For example, let us define the boundary contribution of non-geometric terms,
\begin{equation} \Delta(M_{\mathrm{LC}}+M_{\mathrm{IC}})=(M_{\mathrm{LC}}+M_{\mathrm{IC}})_{t}-(M_{\mathrm{LC}}+M_{\mathrm{IC}})_{b}
\end{equation}
where the subscripts $t$ and $b$ denote ``whole sample'' and ``bulk region'', respectively.
Their convergence processes are plotted in the last row of Fig. \ref{FigM-c=1}. For comparison,
the term $M_{\mathrm{BC}}$ averaged over the whole sample and that over the bulk region are also plotted.
It is interesting to notice that the magnitude of $\Delta(M_{\mathrm{LC}}+M_{\mathrm{IC}})$ converges to the finite value of $M_{\mathrm{BC}}$ averaged over the bulk region in Fig. \ref{FigM-c=1}(d). Remember that $M_{\mathrm{BC}}^{\mathrm{bulk}}=-M_{\mathrm{BC}}^{\mathrm{boundary}}$\cite{Bianco2013}. Therefore in other words, including the boundary allows $M_{\mathrm{LC}}$ and $M_{\mathrm{IC}}$ to complement the loss of $M_{\mathrm{BC}}$ averaged over the boundary, which we call it the compensation effect. It similarly makes sense for the system without e-h symmetry. In Fig. \ref{FigM-c=1}(g), owing to e-h symmetry broken, $M_{\mathrm{LC}}$ and $M_{\mathrm{IC}}$ are no longer opposite, even for a bulk average. Along with the influence of the boundary, the asymmetry between them is further enlarged. Nevertheless, the magnitude of $\Delta(M_{\mathrm{LC}}+M_{\mathrm{IC}})$ still converges to the nonzero value of $M_{\mathrm{BC}}$ with increasing sample size $N$ in Fig. \ref{FigM-c=1}(h). Therefore, it make sure that the total OM calculated by the whole sample average method in $r$ space [Eq. (\ref{EqMr-local})] is consistent with the $k$ space method Eq. (\ref{EqMk}).

We plot the total OM $M$ as a function of the Fermi energy $\mu$ in Fig. \ref{FigM-k&r(c=1)}, calculated from the $k$ space method (black line) and the whole sample average method (red line). For all $\mu$, and independent of e-h symmetry, they match perfectly. This effectively verifies the correctness of including the boundary. As for the magnitude of OM, an important observation is that its maximum values sharply appear at the top of the valence band and the bottom of the conduction band, where the topological edge states end. By comparing with the case of $C=0$ in Fig. \ref{FigM-EF} (c) and (f), we can see that now the maximum OM has been increased remarkably. This suggests that nontrivial topology is still an efficient means to enhance OM. Similar phenomenon has been observed in concrete materials\cite{XiDai2019,JPHanke2016,Lux2018,Helin2020}.

\begin{figure}[htbp]
	\includegraphics*[width=0.48\textwidth]{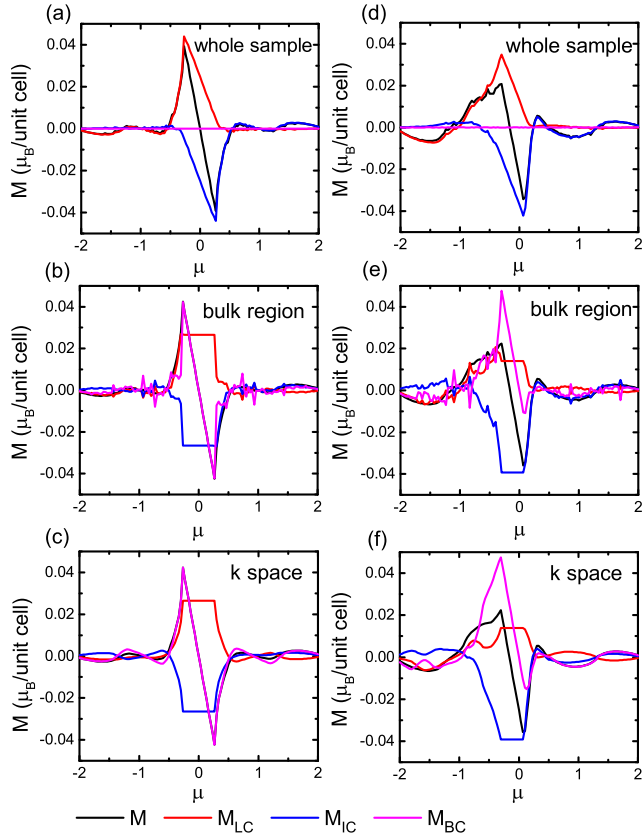}
	\caption{ The OM $M$ and its constituents: $M_{\mathrm{LC}}$, $M_{\mathrm{IC}}$ ,$M_{\mathrm{BC}}$, as functions of the Fermi energy $\mu$ for the QAH bilayer with $C=1$. Upper row: average over the whole sample. Middle row: average over the bulk. Lower row: $k$ space method. The left (right) column corresponds to the case with (without) e-h symmetry. Model parameters are identical to Fig. \ref{FigM-c=1}. }
	\label{FigM-EF(c=1)}
\end{figure}

For details, three terms as functions of the Fermi energy $\mu$ are further investigated, by using the $k$ space method, the bulk average and the whole sample average in $r$ space. The results are shown in Fig. \ref{FigM-EF(c=1)}. We firstly focus on the left row, associated with the e-h symmetric case. From the comparison of Fig. \ref{FigM-EF(c=1)}(b) and (c), we can see that for each term, the result from the bulk average is consistent with that from the $k$ space method. For the whole sample average, although the total OM still matches (Fig. \ref{FigM-k&r(c=1)}), such a term-to-term correspondence with $k$ space method does not hold any more [Fig. \ref{FigM-EF(c=1)}(a) and (c)]. We will check what happens to each constituent term.

First, let us focus on the the bulk gap region, $\mu\in(-0.3,0.3)$. For the bulk average, $M_{\mathrm{LC}}$ and $M_{\mathrm{IC}}$ remain unchanged with varying $\mu$, while $M_{\mathrm{BC}}$ changes linearly. The slope satisfies $\frac{\partial M}{\partial \mu}=\frac{\partial M_{\mathrm{BC}}}{\partial \mu}\propto C$ which can be deduced from Eqs. (\ref{EqMk}) or (\ref{EqMr-local}). When we switch to the whole sample average as shown in Fig. \ref{FigM-EF(c=1)}(a), on the other hand, $M_{\mathrm{BC}}$ is always $0$ when $\mu$ is within the bulk gap. This is consistent with the previous work\cite{Bianco2013}. Surprisingly, within this energy region $M_{\mathrm{LC}}$ and $M_{\mathrm{IC}}$ change linearly with $\mu$, instead of remaining constant in Fig. \ref{FigM-EF(c=1)} (b). The joint effect of them results in a linear relationship between $M$ and $\mu$, and $\frac{\partial M}{\partial \mu}\propto C$ still holds. Such a story is similarly applicable to the system with broken e-h symmetry as presented in Fig. \ref{FigM-EF(c=1)}(d)-\ref{FigM-EF(c=1)}(f). This further verifies the above conclusion that the compensation effect of the boundary ensures $M_{\mathrm{LC}}$ and $M_{\mathrm{IC}}$ to supplement the loss of $M_{\mathrm{BC}}$, when the real space average is switched from bulk region to whole sample. Thus, it is also reliable to average over the whole sample for $r$ space method even for the case of $C=1$.

\subsection{B. Chern number $C=2$}

\begin{figure}[htbp]
	\includegraphics*[width=0.5\textwidth]{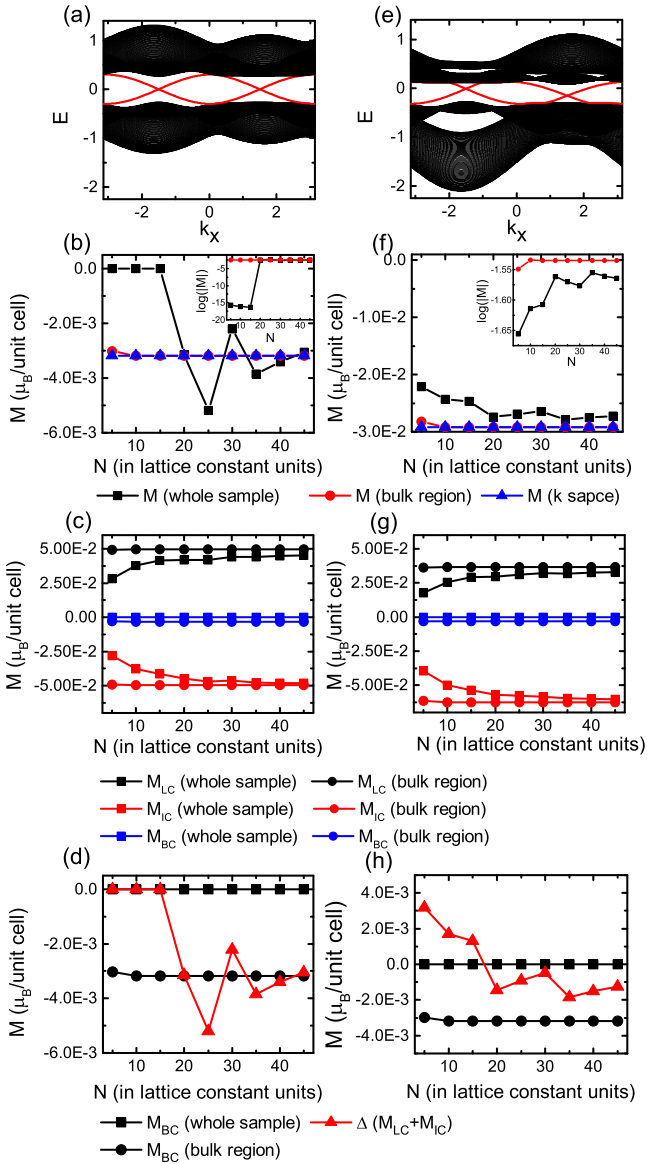}
	\caption{ Similar to Fig. \ref{FigM-c=1} but for the bilayer with $C=2$. First row: band structures of a ribbon. Second row: the total OM as a function of the sample size $N$ calculated from three methods. Third row: $M_{\mathrm{LC}}$ ,$M_{\mathrm{IC}}$ and $M_{\mathrm{BC}}$ as functions of the sample size $N$. Fourth row: comparison of $M_{\mathrm{BC}}$ with $\Delta(M_{\mathrm{LC}}+M_{\mathrm{IC}})$ as functions of the sample size $N$. The left (right) column corresponds to the case with (without) e-h symmetry. $M_2=0.5$, and other model parameters are the same as Fig. \ref{FigM-c=1}. }
	\label{FigM-c=2}
\end{figure}

\begin{figure}[htbp]
	\includegraphics*[width=0.48\textwidth,bb=38 25 496 223]{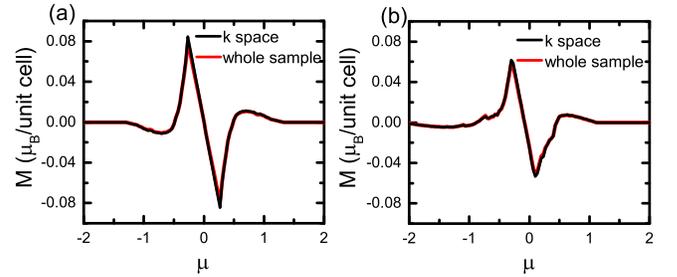}
	\caption{ Similar to Fig. \ref{FigM-k&r(c=1)} but for the bilayer with $C=2$, total OM as a function of the Fermi energy $\mu$, calculated from two methods. (a) With e-h symmetry. (b) Without e-h symmetry. Model parameters are identical to Fig. \ref{FigM-c=2}.}
	\label{FigM-k&r(c=2)}
\end{figure}

The above conclusions can be generalized to the case with $C=2$. Similar to Fig. \ref{FigM-c=1}, in Fig. \ref{FigM-c=2} we present the size convergence of total OM from different methods (second row), three constituent terms (third row) and the contribution from the boundary (fourth row). Similar to the case of $C=1$, $M$ calculated from the whole sample method fluctuates more than that from other methods. However, it still converges to the correct value for sufficiently large sample. As a matter of fact, as presented in Fig. \ref{FigM-k&r(c=2)}, the total OM from the whole sample average (red line) is perfectly consistent with that from the $k$ space method (black line) for all Fermi energies $\mu$.

\begin{figure}[htbp]
	\includegraphics*[width=0.48\textwidth]{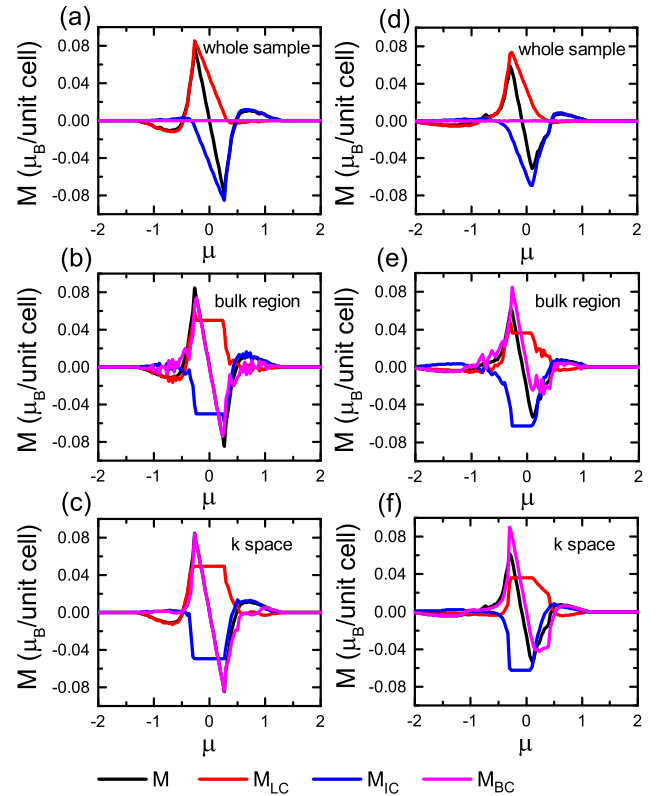}
	\caption{ Similar to Fig. \ref{FigM-EF(c=1)} but for the bilayer with $C=2$. The OM and its constituents as functions of the Fermi energy $\mu$. Upper row: average over the whole sample. Middle row: average over the bulk. Lower row: $k$ space method. The left (right) column corresponds to the case with (without) e-h symmetry. Model parameters are identical to Fig. \ref{FigM-c=2}.}
	\label{FigM-EF(c=2)}
\end{figure}

Also similar to the $C=1$ case, the inclusion of boundary affects all three components: $M_{\mathrm{LC}}$, $M_{\mathrm{IC}}$ and $M_{\mathrm{BC}}$. If the system has e-h symmetry, $M_{\mathrm{LC}}$ (black circle line) and $M_{\mathrm{IC}}$ (red circle line) are symmetric with respect to zero in Fig. \ref{FigM-c=2}(c) when averaged over bulk, while they are no longer symmetric when averaged over the whole sample. Independent of e-h symmetry, the compensation effect we met in the case of $C=1$ does work here.
For example, $\Delta(M_{\mathrm{LC}}+M_{\mathrm{IC}})$ still converges to the finite value of $M_{\mathrm{BC}}$ averaged over the bulk region as shown in the last row of Fig. \ref{FigM-c=2}. Furthermore, such a compensation effect works for all $\mu$ values, including the bulk gap with large orbital magnetic moment from edge states, as presented in Fig. \ref{FigM-EF(c=2)}.

One more thing should also be noticed by comparing the vertical scales in Fig. \ref{FigM-k&r(c=2)} and in Fig. \ref{FigM-k&r(c=1)}. The peak value of the total OM with $C=2$ is even larger than that in the case with $C=1$. This again confirms that nontrivial topology, i.e., Chern number, or at least remarkably nonzero integrated Berry curvature, is one of the most effective factor to enhance OM\cite{JPHanke2016,Lux2018}.

\section{VI. Size Effects}
In this section, we discuss some size effects of the real space algorithm.
In previous studies, when discussing the roles of boundaries, we separated the bulk and boundary regions in a manner of a fixed fraction: the central quarter of the sample is adopted as the bulk, and the rest part is the boundary. Now we discuss the effect of different manners of separation by varying the thickness of the boundary, $L_{B}$. In Fig. \ref{lattice}, bulk averaged OMs with different definitions of $L_{B}$ are plotted, for the cases of $C=1$ (left column) and $C=2$ (right column) respectively. Notice the $L_{B}=0$ case corresponds to the whole sample average. Here, the values of constituent terms may fluctuate with different $L_{B}$ but the total OM is invariant and consistent with that from the $k$ space method. In other words, for a sufficiently large sample, the details of separating the bulk and boundary regions are not important, and again, even the whole sample average is also valid.

\begin{figure}[htbp]
	\includegraphics*[width=0.48\textwidth]{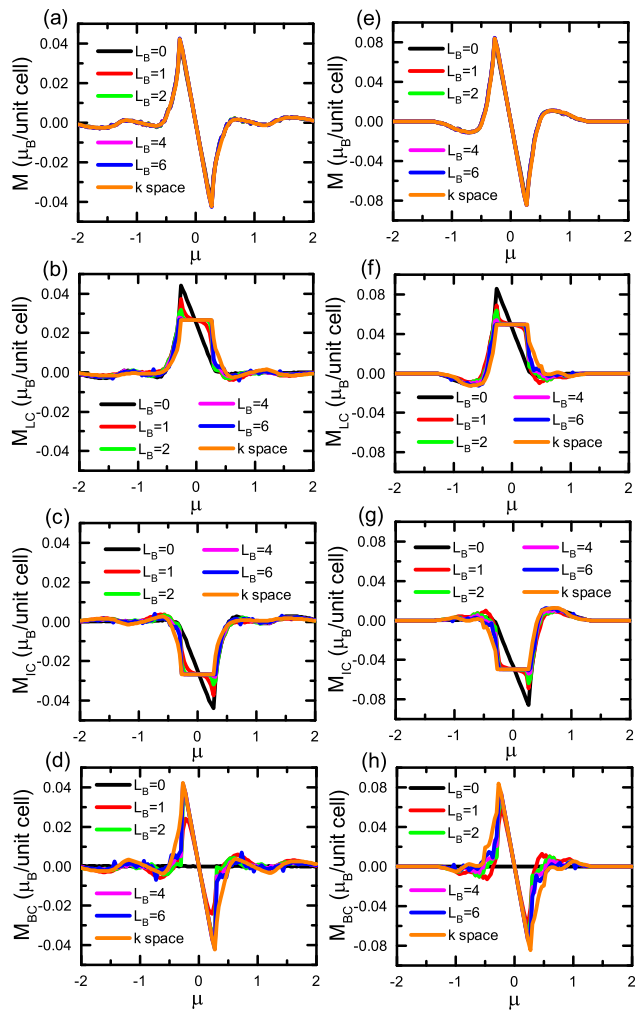}
	\caption{ The bulk averaged OM and its constituents as functions of the Fermi energy $\mu$ with different adoptions of boundary thickness $L_{B}$. The left (right) column denotes the bilayer with $C=1$ ($C=2$). First row: the bulk average of $M$. Second row: the bulk average of $M_{\mathrm{LC}}$. Third row: the bulk average of $M_{\mathrm{IC}}$. Fourth row: the bulk average of $M_{\mathrm{BC}}$. Different colors represent different thickness of boundaries. Orange lines signify the results of $k$ space method. The sample size $N=40$, and other model parameters are identical to Fig. \ref{FigM-c=1}.}
	\label{Boundary}
\end{figure}

\begin{figure}[htbp]
	\includegraphics*[width=0.48\textwidth]{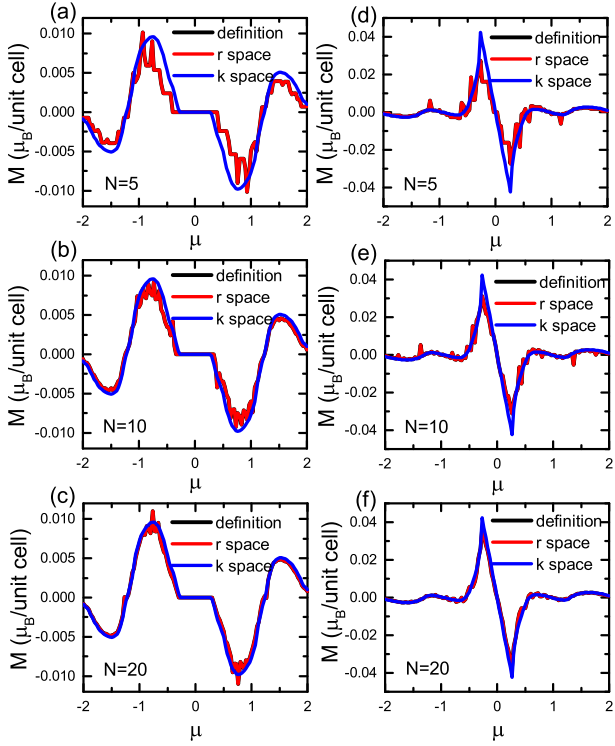}
	\caption{ The total OM as a function of the Fermi energy $\mu$ for very small samples. The left (right) column represent the sample size $N=5$ ($N=10$). Upper row: the bilayer with $C=0$.
	Middle row: the bilayer with $C=1$. Lower row: the bilayer with $C=2$. Black lines: the OM calculated from the definition Eq. (\ref{EqMFirstPrinciple}). Red lines: the OM calculated from $r$ space method Eq. (\ref{EqMr-local}). Blue lines: the OM calculated from $k$ space method Eq. (\ref{EqMk}). In all panels, black curves have been covered by red curves due to perfect coincidence between them. Model parameters are identical to Fig. \ref{FigM-c=1}. }
	\label{smallsize}
\end{figure}

The focus so far is the thermodynamic limit, i.e., real space results for a sufficiently large sample. With the progress of nanostructure fabrication technologies in thin films, clusters and manmade structures, it is also practical to examine the OM of very small samples. To check the validity of current real space algorithm in this case, we calculate the finite-size OM from the ``first principle'' expression\cite{Bianco2013,Thonhauser2005}
\begin{equation}
	M=-\frac{e}{2cA}\sum_{i}\langle\varphi_{i}|\bm{r}\times\bm{v}|\varphi_{i}\rangle,
	\label{EqMFirstPrinciple}
\end{equation}
and then compare it with those from $r$ space (whole-sample average) and $k$ space method. Since the size is very small (less than 100 sites), the separation of bulk and boundaries makes no sense, and therefore the $r$ space algorithm is just a whole-sample average. The data are shown in Fig. \ref{smallsize}. It is interesting to see the perfect consistency between the results from Eq. (\ref{EqMFirstPrinciple}) (black line) and the whole-sample average (red line). This example again confirms the importance of this whole-sample average in real space calculations.

\section{VII. Summary}
In summary, based on a bilayer model with an adjustable Chern number, we numerically investigated the role of boundaries in the real space algorithm of calculating orbital magnetization $M$, by scrutinizing its constituent terms $M_{\mathrm{LC}}$, $M_{\mathrm{IC}}$, $M_{\mathrm{BC}}$, as well as two different averages: over the bulk or over the whole sample. The first key finding is that for the total OM, both average methods converge to the same value calculated from $k$ space method, for sufficiently large sample size. This conclusion does not depend on the topological nature of the system, i.e., Chern number at half filling.

The boundary region has a significant contribution to the geometric term $M_{\mathrm{BC}}$, especially in the case of nontrivial topology. However, it also has a non-negligible contribution to non-geometric terms $M_{\mathrm{LC}}$ and $M_{\mathrm{IC}}$, and
these two types of contributions compensate each other exactly, which results in the consistency of two average methods in the real space.
The compensation effect is further manifested by the property that $M_{\mathrm{LC}}$ and $M_{\mathrm{IC}}$ inherit the linear relationship between $M_{\mathrm{BC}}$ and the Fermi energy $\mu$ for Chern insulators when $\mu$ is in the bulk band gap. This is different from the previous conclusion which advocates that the boundaries of Chern insulators should not be considered in $r$ space expression because of the questionable contribution of $M_{\mathrm{BC}}$ from boundaries. From the comparison among systems with different Chern numbers, the systems with high Chern number are suitable candidate materials with strong orbital magnetism. This provides a theoretical reference for finding ways to enhance orbital magnetism. This $r$ space method of a whole-sample average is also important for nanostructures, where the separation between the bulk and boundaries seems technologically problematic. These clarifications will also be instructive to further studies on the orbital magnetoelectric effects in real space.

\section{Acknowledgements}
This work was supported by National Natural Science
Foundation of China under Grant Nos. 12104108, 11774336, and 12147112,
and the Starting Research Fund from
Guangzhou University under Grant Nos. RQ2020082 and 62104360.

\section{Appendix: Calculation of Orbital Magnetization}
\setcounter{equation}{0}

\renewcommand{\theequation}{A\arabic{equation}}

For a two dimensional material, the OM $M$ is defined as the orbital moment per unit area\cite{Ceresoli2006},
\begin{equation}
	M=-\frac{e}{2cA}\sum_{i}\langle\varphi_{i}|\bm{r}\times\bm{v}|\varphi_{i}\rangle
	\label{EqMApp}
\end{equation}
where $e$ is the magnitude of the electronic charge, $c$ is the vacuum speed of light and $A$ is the sample area. $|\varphi_{i}\rangle$ are the occupied eigenstates, $\bm{r}$ is the position operator and $\bm{v}$ is the velocity operator which satisfies
\begin{equation}
	v=-\frac{i}{\hbar}\left[ \bm{r},H\right]
	\label{EqvApp}
\end{equation}
Here, $H$ is the Hamiltonian operator. It is obvious that the above expression can be directly applied to open boundary conditions(OBCs). Due to the non-periodic nature of the position operator $r$, it requires further derivation to obtain $k$ space expression for periodic boundary conditions(PBCs) which is known as the modern theory of orbital magnetization.

For a two dimensional gapped insulator, electrons are in the $x$-$y$ plane and the OM $M$ is along $z$ direction. Eq. (\ref{EqMApp}) can be rewritten in the $r$ space as follows,
\begin{equation}\label{EqMrApp}
\begin{aligned}
	&M=M_{\mathrm{LC}}+M_{\mathrm{IC}}\\
    &M_{\mathrm{LC}}=\frac{e}{\hbar cA}\mathrm{ImTr}\left\lbrace PxQHQyP\right\rbrace \\
	&M_{\mathrm{IC}}=-\frac{e}{\hbar cA}\mathrm{ImTr}\left\lbrace QxPHPyQ\right\rbrace \\
\end{aligned}
\end{equation}
where $M_{\mathrm{LC}}$ and $M_{\mathrm{IC}}$ correspond to the local circulation (LC), the itinerant circulation (IC) respectively. The ground-state projector $P$ and its complement $Q$ satisfy
\begin{equation}\label{EqPApp}
	\begin{aligned}
		&P=\sum_{\varepsilon_{i}\leqslant \mu}|\varphi_{i}\rangle \langle \varphi_{i}|
		&Q=1-P\\	
	\end{aligned}
\end{equation}
where $\mu$ is the Fermi energy, and $\varepsilon_{i}$ is the eigenvalue evaluated by $H({\bm{r}})|\varphi_{i}\rangle=\varepsilon_{i}|\varphi_{i}\rangle$. The position operator $r$ becomes well-defined when "sandwiched" between a $P$ and a $Q$ within PBCs. Generalizing equation (\ref{EqMrApp}) to Chern insulators and metals, there is one more component coming from the Berry curvature(BC) contributions. We call it $M_{\mathrm{BC}}$, which is associated with the topological properties of materials\cite{Wang2020}. Hence,
\begin{equation}\label{EqMr-chernApp}
	\begin{aligned}
		&M=M_{\mathrm{LC}}+M_{\mathrm{IC}}+M_{\mathrm{BC}}\\
		&M_{\mathrm{LC}}=\frac{e}{\hbar cA}\mathrm{ImTr}\left\lbrace PxQHQyP\right\rbrace \\
		&M_{\mathrm{IC}}=-\frac{e}{\hbar cA}\mathrm{ImTr}\left\lbrace QxPHPyQ\right\rbrace \\
        &M_{\mathrm{BC}}=-\mu\frac{e}{2\pi\hbar cA} \mathcal{C}\\
		&\mathcal{C}=4\pi \mathrm{ImTr}\left\lbrace QxPyQ\right\rbrace
	\end{aligned}
\end{equation}
In addition, the $r$ space expression also gives the local description of $M$
\begin{equation}\label{EqMr-localApp}
	\begin{aligned}
		&M=\frac{1}{A}\int d{\bm{r}}\mathit{m}\left( \bm{r}\right) \\
		&{\mathit{m}\left( \bm{r}\right)}=m_{\mathrm{LC}}\left( \bm{r}\right)+m_{\mathrm{IC}}\left( \bm{r}\right)+m_{\mathrm{BC}}\left( \bm{r}\right)\\
		&{m_{\mathrm{LC}}\left( \bm{r}\right)}=\frac{e}{\hbar c}\mathrm{Im}\langle \bm{r}| PxQHQyP|\bm{r}\rangle\\
	    &{m_{\mathrm{IC}}\left( \bm{r}\right)}=-\frac{e}{\hbar c}\mathrm{Im}\langle \bm{r}| QxPHPyQ|\bm{r}\rangle \\
        &{m_{\mathrm{BC}}\left( \bm{r}\right)}=-\mu\frac{e}{2\pi\hbar\ c}\emph{c}\left( \bm{r}\right) \\
		&{\emph{c}\left( \bm{r}\right)}=4\pi \mathrm{Im}\langle \bm{r}|QxPyQ|\bm{r}\rangle \\
	\end{aligned}
\end{equation}
where $m_{\mathrm{LC}}\left( \bm{r}\right)$, $m_{\mathrm{IC}}\left( \bm{r}\right)$ and $m_{\mathrm{BC}}\left( \bm{r}\right)$ correspond to the local properties of the three terms in Eq.(\ref{EqMr-chernApp}). $\emph{c}\left( \bm{r}\right)$ is called the topological marker.

Using the relation between Wannier representation and Bloch representation, Eq.(\ref{EqMr-chernApp}) can be linked with $k$ space expression which is called the modern theory of OM for a Chern insulator\cite{Ceresoli2006,Malashevich2010}. In the $k$ space, the $M$ is expressed as a Brillouin-zone integral of Bloch-orbital matrix elements\cite{Resta2010,Malashevich2010,Ceresoli2006}. The macroscopic OM of a general two dimensional crystalline solid is
 \begin{equation}\label{EqMkApp}
\begin{aligned}
 &M=M_{\mathrm{LC}}+M_{\mathrm{IC}}+M_{\mathrm{BC}}\\
 &M_{\mathrm{LC}}=\frac{e}{2\hbar c}\mathrm{Im}\sum_n\int_{\varepsilon_{nk}\leqslant\mu}
 \frac{\mathrm{d}\bm{k}}{(2\pi)^2}\langle\partial_{\bm{k}}u_{n\bm{k}}|
 \times H_{\bm{k}} |\partial_{\bm{k}}u_{n\bm{k}}\rangle \\
 &M_{\mathrm{IC}}=\frac{e}{2\hbar c}\mathrm{Im}\sum_n\int_{\varepsilon_{nk}\leqslant\mu}
 \frac{\mathrm{d}\bm{k}}{(2\pi)^2}\varepsilon_{n\bm{k}}\langle\partial_{\bm{k}}u_{n\bm{k}}|
 \times|\partial_{\bm{k}}u_{n\bm{k}}\rangle \\
 &M_{\mathrm{BC}}=-\mu \frac{e}{2\pi\hbar c}\mathcal{C}\\
 &\mathcal{C}=\frac{1}{2\pi}\mathrm{Im}\sum_n\int_{\varepsilon_{nk}\leqslant\mu}
 \mathrm{d}\bm{k}\langle\partial_{\bm{k}}u_{n\bm{k}}|
 \times|\partial_{\bm{k}}u_{n\bm{k}}\rangle
\end{aligned}
\end{equation}
where $|u_{n\bm{k}}\rangle$ is the cell-periodic Bloch function, and $\varepsilon_{nk}$ is the Bloch eigenvalue which satisfies $H_{\bm{k}}|u_{n\bm{k}}\rangle=\varepsilon_{nk}|u_{n\bm{k}}\rangle$. All summations in equation (\ref{EqMkApp}) are over occupied bands $n$ up to $\mu$.
In numerical calculations, the derivatives $|{\partial}_{\bm{k\alpha}} u_{n\bm{k}}\rangle$ are evaluated on a mesh of discretized Brillouin zone. Since a simple finite-deference of the wave functions $|u_{n\bm{k}}\rangle$ cannot ensure a fixed gauge on neighboring grid points, we adopt the discretized covariant derivative $|\tilde{\partial}_{\bm{k\alpha}} u_{n\bm{k}}\rangle$ which involves linear combination of occupied states and a local fixed gauge around a definite grid point.
The definition of $|\tilde{\partial}_{\bm{k\alpha}} u_{n\bm{k}}\rangle$ is
\begin{equation}
	{|\tilde{\partial}_{\bm{k\alpha}}u_{n\bm{k}}\rangle}=Q_{\bm{k}}|\partial_{\bm{k\alpha}} u_{n\bm{k}}\rangle
	\label{u_nkApp}
\end{equation}
where $Q_{\bm{k}}$ is the Fourier transformation of Eq.(\ref{EqPApp}). For more details see appendix A of reference \cite{Ceresoli2006}. Let us define\cite{Ceresoli2006,Souza2008}
\begin{equation}\label{EqfkApp}
	\begin{aligned}
		&\tilde{f}_{\bm{k},i}=\frac{1}{v}\in_{ijl}q_{i}\sum_n\mathrm{Im}\langle\tilde{\partial}_{\bm{kj}}u_{n\bm{k}}\
		|\tilde{\partial}_{\bm{kl}}u_{n\bm{k}}\rangle \\
		&\tilde{g}_{\bm{k},i}=\frac{1}{v}\in_{ijl}q_{i}\sum_n\mathrm{Im}\langle\tilde{\partial}_{\bm{kj}}u_{n\bm{k}}\
	  \mathrm{H}_{\bm{k}}|\tilde{\partial}_{\bm{kl}}u_{n\bm{k}}\rangle \\
		&\tilde{h}_{\bm{k},i}=\frac{1}{v}\in_{ijl}q_{i}\sum_n\varepsilon_{nm\bm{k}}\mathrm{Im}\langle\tilde{\partial}_{\bm{kj}}u_{m\bm{k}}\
		|\tilde{\partial}_{\bm{kl}}u_{n\bm{k}}\rangle
	\end{aligned}
\end{equation}
where $q_{i}$ denotes the primitive reciprocal vectors that define the $k$ mesh along the $i$th direction, and $v$ is the volume of the unit cell of the $k$ mesh. Thus, we can further rewrite Eq.(\ref{EqMkApp}) as follows
 \begin{equation}\label{EqMk2App}
	\begin{aligned}
		&M=\tilde{M}_{\mathrm{LC}}+\tilde{M}_{\mathrm{IC}}+M_{\mathrm{BC}}\\
		&\tilde{M}_{\mathrm{LC}}=\frac{-e}{2\hbar c}\int_{\mathrm{BZ}}
		\frac{\mathrm{d}\bm{k}}{(2\pi)^2}\tilde{g}_{\bm{k}}\\
		&\tilde{M}_{\mathrm{IC}}=\frac{-e}{2\hbar c}\int_{\mathrm{BZ}}
		\frac{\mathrm{d}\bm{k}}{(2\pi)^2}\tilde{h}_{\bm{k}}\\
        &M_{\mathrm{BC}}=-\mu \frac{e}{2\pi\hbar c}\mathcal{C}\\
		&\mathcal{C}=\frac{1}{2\pi}\int_{\mathrm{BZ}}\tilde{f}_{\bm{k}}
	\end{aligned}
\end{equation}
In this manuscript, we have adopted both the $k$ space and $r$ space expression to study OMs. We will focus on the influence of the boundaries on the $M$ and its three components: $M_{\mathrm{LC}}$, $M_{\mathrm{IC}}$ and $M_{\mathrm{BC}}$.

\end{document}